\documentclass{article}

\usepackage{arxiv}
\usepackage{makeidx}
\usepackage{graphicx}
\usepackage{amsmath,amssymb} 
\usepackage{color}
\usepackage[utf8]{inputenc} 
\usepackage[T1]{fontenc}    
\usepackage{hyperref}       
\usepackage{url}            
\usepackage{booktabs}       
\usepackage{amsfonts}       
\usepackage{nicefrac}       
\usepackage{microtype}      
\usepackage{subcaption}
\usepackage{listings}
\usepackage{appendix}

\AtBeginDocument{%
  \providecommand\BibTeX{{%
    \normalfont B\kern-0.5em{\scshape i\kern-0.25em b}\kern-0.8em\TeX}}}

\begin{document}

\title{Defending Against Adversarial Denial-of-Service Data Poisoning Attacks}

\author{
  Nicolas M. M\"uller\thanks{Equal Contribution}\\
  Fraunhofer AISEC \\
  Garching near Munich, Germany \\
  \texttt{nicolas.mueller@aisec.fraunhofer.de}
   \And
  Simon Roschmann\footnotemark[1]\\
  Technical University of Munich \\
  Garching near Munich, Germany \\
  \texttt{simon.roschmann@tum.de}
   \And
   Konstantin B\"ottinger \\
  Fraunhofer AISEC \\
  Garching near Munich, Germany \\
  \texttt{konstantin.boettinger@aisec.fraunhofer.de}
}
\maketitle



\begin{abstract}
Data poisoning is one of the most relevant security threats against machine learning and data-driven technologies. 
Since many applications rely on untrusted training data, an attacker can easily craft malicious samples and inject them into the training dataset to degrade the performance of machine learning models. 
As recent work has shown, such Denial-of-Service (DoS) data poisoning attacks are highly effective.

To mitigate this threat, we propose a new approach of detecting DoS poisoned instances.
In comparison to related work, we deviate from clustering and anomaly detection based approaches, which often suffer from the curse of dimensionality and arbitrary anomaly threshold selection.
Rather, our defence is based on extracting information from the training data in such a generalized manner that we can identify poisoned samples based on the information present in the unpoisoned portion of the data.
We evaluate our defence against two DoS poisoning attacks and seven datasets, and find that it reliably identifies poisoned instances.
In comparison to related work, our defence improves false positive / false negative rates by at least 50\%, often more.
\end{abstract}





\section{Introduction}\label{s:introduction}
In recent years, machine learning has led to new opportunities and breakthroughs in the fields of image classification \cite{he2016deep,zoph2018learning} and speech recognition \cite{graves2013speech,xiong2016achieving}.
Machine learning has also been applied in security-sensitive applications including face  recognition \cite{sun2014deep}, fingerprint identification \cite{wang2014fingerprint}, autonomous driving \cite{huval2015empirical} and the detection of spam \cite{nelson2008exploiting}, malware \cite{saxe2015deep} and network intrusion \cite{javaid2016deep}.

However, machine learning models are vulnerable and can be targeted by attackers. Huang et al. \cite{huang2011adversarial} have created a taxonomy for adversarial machine learning that distinguishes between causative (poisoning) attacks compromising the learning mechanism at train time and exploratory (evasion) attacks circumventing the learning mechanism at test time.
Evasion attacks exploit blind spots in the classifier that allow malicious activities to remain undetected, e.g. test samples are manipulated to be misclassified at run time. 
Biggio et al. \cite{biggio2013evasion} show how changing malicious \emph{pdf} files can make them look legitimate and prevent detection by the classifier.

In contrast, poisoning attacks manipulate training data to compromise the integrity or availability of a classifier. 
DoS poisoning attacks aim at maximally degrading the performance of a model to make an application or service unusable. 
For example, Nelson et al. \cite{nelson2008exploiting} use DoS data poisoning to attack a spam filter.
By sending attack emails, which contain many words likely to occur in legitimate emails, they increase the likelihood of future benign emails to be marked as spam by the learning algorithm.
As a spam filter relies on very low false-positive rates, it is effectively shut down by the attack.
Further poisoning attacks are explored in \cite{biggio2011support,biggio2012poisoning,koh2017understanding,munoz2017towards,xiao2015feature} and considered to be one of the most significant threats for systems that rely upon collecting untrusted data.

To mitigate the threat of DoS poisoning, previous work has developed defences based on measuring the influence of samples on a model \cite{koh2017understanding,nelson2008exploiting} or on clustering and anomaly detection \cite{paudice2018detection,paudice2018label}. 
However, most defences are computationally inefficient or susceptible to the curse of dimensionality \cite{Keogh2017}.
Also, anomaly detection based approaches often suffer from arbitrary selection of detection thresholds.

\textbf{Contribution.}
The contribution of this paper is a new approach to defending against strong DoS poisoning attacks, such as poisoning attacks with back-gradient optimization presented in \cite{munoz2017towards}.
Our defence is based on the re-evaluation of training samples by the poisoned classifier, which leverages the information present in the unpoisoned part of the dataset to identify and iteratively remove the poisoned instances.
No interaction with a human oracle is required.
This approach overcomes difficulties and constraints of previously proposed defences. 
Compared to outlier detection, our approach is not susceptible to the curse of dimensionality. 
Moreover, we do not put any constraints on the datasets, do not require knowledge about the fraction of poison samples within our training dataset and do not restrict our defence to linear or binary models.
An experimental evaluation of two poisoning attacks against the classification of seven datasets with neural networks shows the effectiveness of our proposed defence methodology.
In comparison to related work, our defence improves false positive / false negative rates by at least 50\%, often more.

\section{Related Work}\label{s:related_work}

\subsection{Poisoning Attack}\label{ss:poisoning_attack}

A simple label flipping poisoning attack against the binary classification with Support Vector Machines is proposed in \cite{biggio2011support}.
To maximize the classification error, the attacker chooses a subset of influential poison samples by evaluating the impact of samples with flipped labels on a separate validation set.
Biggio et al. \cite{biggio2012poisoning} have been the first to demonstrate a poisoning attack strategy against Support Vector Machines by solving a bi-level optimization problem (see Equation~\ref{eq:1},~\ref{eq:2}) using the Karush-Kuhn-Tucker (KKT) conditions. 
Xiao et al. \cite{xiao2015feature} apply the same approach to poison Ridge regression, Lasso regression and Elastic Net regression.
Influence functions are discovered in \cite{koh2017understanding} to measure the impact that a training sample has on the classifier's loss without performing computationally expensive retraining. 
In the context of data poisoning this approach can be used to identify samples that should be modified to maximally compromise the classifier's performance.

While all these approaches have focused on binary classification, Muñoz-González et al. \cite{munoz2017towards} target gradient-based deep learning algorithms for multiclass classification. 
They extend the commonly used threat model by introducing the concept of error specificity and overcome limitations regarding computation, memory and accuracy of poisoning attacks based on KKT conditions. 

\subsection{Poisoning Defence}\label{ss:poisoning_defence}
To mitigate the threat of DoS poisoning attacks against spam filters, Nelson et al. \cite{nelson2008exploiting} propose to measure the impact of each training sample on the classifier's performance. This defence can effectively determine samples that decrease the accuracy and should be discarded. 
However, it is computationally expensive 
and can suffer from overfitting when the dataset is small compared to the number of features.
A similar approach based on influence functions is presented in \cite{koh2017understanding}. 

The authors in \cite{feng2014robust} propose a two step defence strategy for logistic regression and classification. 
After applying outlier detection, the algorithm is trained solving an optimization problem based on the sorted correlation between the classifier and the remained samples. 
This approach is not feasible in practice as the fraction of poison samples is required for the algorithm to perform well.

There are also different approaches of outlier detection.
The authors of~\cite{paudice2018label} use the k-nearest neighbours algorithm to find instances whose labels differ from that of their neighbours. These instances are considered to be adversarial.
In contrast, the authors of~\cite{paudice2018detection} first split the instances in the poisoned dataset into their respective classes and then use different metrics for outlier detection to identify poison instances with high outlierness score. Note that this defence requires a training set of \emph{trusted} data.
Both approaches highly depend on the selection of appropriate thresholds, and suffer from the curse of dimensionality.

In summary, our defence improves upon related work as follows: 
\begin{itemize}
    \itemsep0em
    \item Our defence  is computationally light, as opposed to previous work.\footnote{See for example the defence presented in \cite{nelson2008exploiting}, which requires model retraining for every single sample.
    Given a dataset with $n=1000$ instances, their algorithm requires fitting a model $1000$ times, while our defence only needs fitting a model twice.}
    \item Our defence is not susceptible to the curse of dimensionality, such as~\cite{paudice2018detection,paudice2018label}.
    \item Our defence does not require the correct percentage of poison samples $\nu$ (as opposed to e.g.~\cite{feng2014robust}), but in fact provides an estimate $\hat{\nu}$.
    \item Our defence keeps FPR and FNR low compared to previous outlier detection (see Table~\ref{tab:defence_comparison}) without the necessity of picking appropriate detection thresholds.
\end{itemize}

\section{Poisoning Attack}\label{s:poisoning_attack}
In this section, we present two DoS poisoning attacks from related work against which we evaluate our defence.

\subsection{Threat Model}\label{ss:threat_model}
Denial-of-Service poisoning attacks aim at violating a system's availability by preventing users from access to its normal functionality. 
The goal of these attacks is making the model inconsistent or unreliable in the target environment \cite{papernot2018sok} (for example to break a spam filter, as shown in \cite{nelson2008exploiting}).


Limited-knowledge attacks simulate a more realistic attack setting, where the training data or learning algorithm is unknown.
In this scenario, an attacker uses substitute datasets and models during poisoning. 
Perfect-knowledge attacks, on the other hand, simulate an attacker that knows everything about the targeted system
and can be used to derive worst-case performance degradation estimates.

Usually, we assume that a dataset is crowd-sourced and that an attacker can introduce a certain percentage $\nu$ of adversarial data points into the training dataset.
In order to create poison samples which are at least somewhat difficult to detect with outlier detection techniques and data pre-filtering, these poisoned samples should not take extreme values (both in feature and target space).

The data available to the attacker is divided into two disjoint sets:
$D_{tr}$ to train the model and $D_{val}$ to sample poison instances $D_p$ and evaluate their impact on test-time data.

\subsection{DoS Poisoning via Label Flipping Attack}\label{ss:flip_attack}

\begin{figure*}[t]
\centering
\begin{subfigure}[b]{0.49\textwidth}
\includegraphics[width=\textwidth]{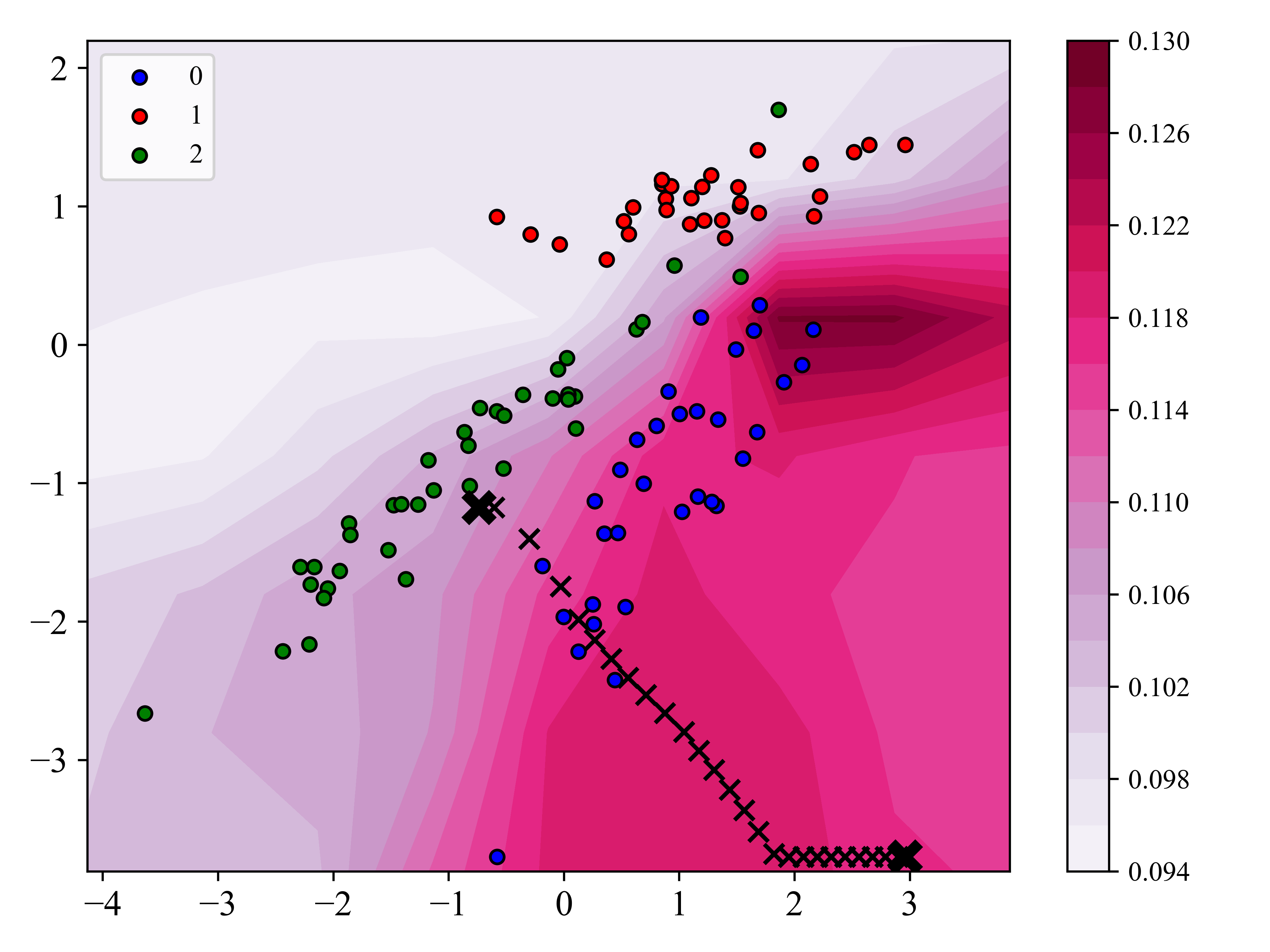}
\caption{Validation loss}
\label{fig:points_heat_map_loss}
\end{subfigure}
\hfill
\begin{subfigure}[b]{0.49\textwidth}
\includegraphics[width=\textwidth]{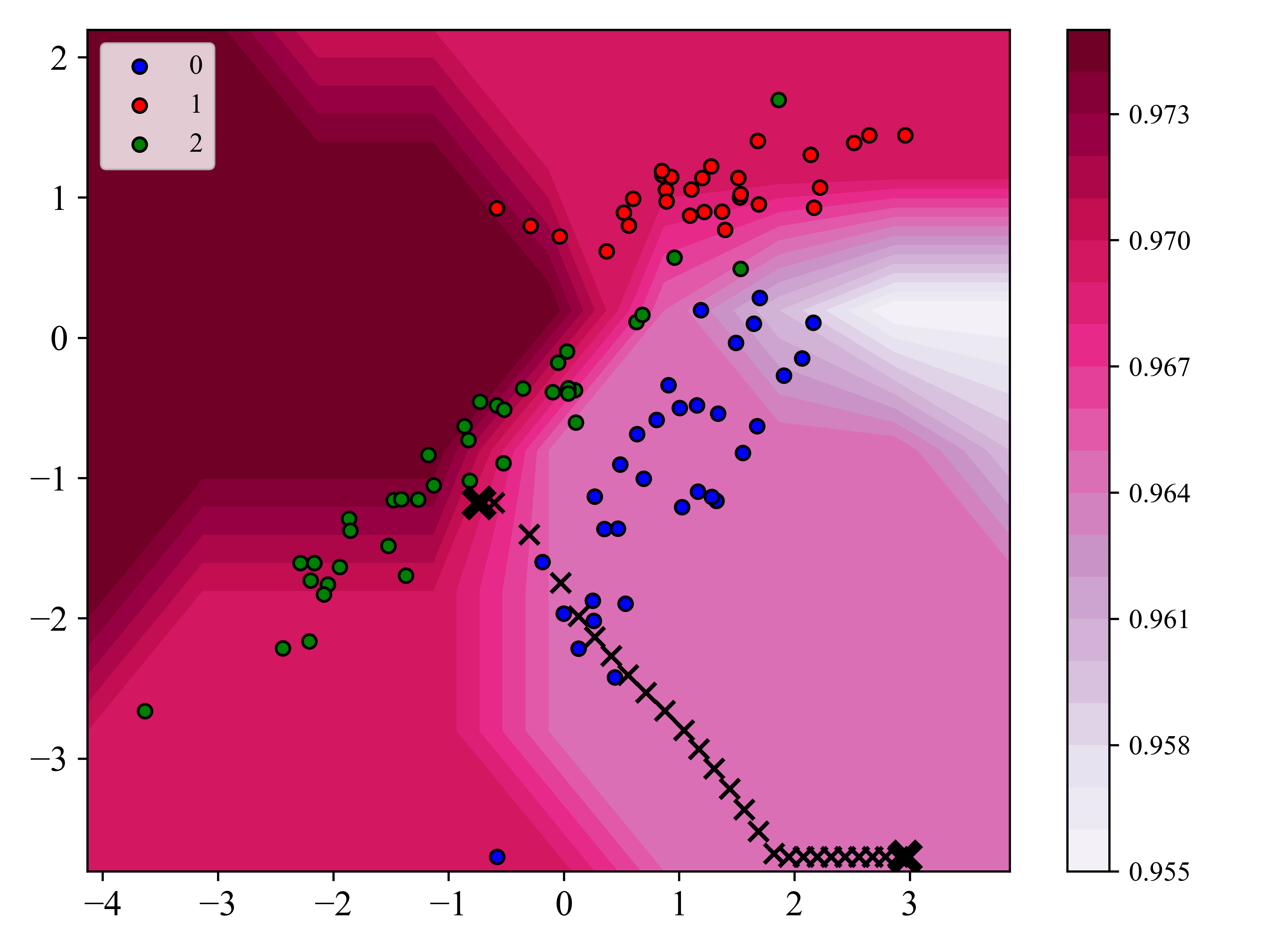}
\caption{Validation accuracy}
\label{fig:points_heat_map_accuracy}
\end{subfigure}
\caption{The trajectory of a malicious sample during a poisoning attack against the two-dimensional $points$ dataset. After initially switching the label from $2$ to $1$, the instance is moved from $x_p=[-0.74, -1.18]$ (region of smaller validation loss) to $x_p=[2.96, -3.7]$ (region of higher validation loss) by iteratively applying back-gradients.}
\label{fig:heat_map}
\end{figure*}

As a baseline attack, we consider Random Label Flipping, where the attacker selects a maximum percentage $\nu$ of instances at random from the training dataset and changes their labels to a class different from the original one \cite{paudice2018label}.
This simple concept of label flipping describes a limited-knowledge attack, as it only requires the attacker to have access to the labels of the training dataset $D_{tr}$.

\subsection{DoS Poisoning via Back-gradient Optimization Attack}\label{ss:bg_attack}
In this section, we briefly describe the current state-of-the art attack for DoS data poisoning \cite{munoz2017towards}.
It is a perfect-knowledge attack that is characterized by iteratively adding poison samples into the training dataset.
An attacker objective function $A$ maps a poisoned model $M$ with parameters $w^*$ and a validation dataset $D_{val}$ to a real number, i.e. $A(D_{val}, w^*) \in \mathbb{R}$.
The attacker aims at finding poisoned instances $D^*_p$ such that $A$ is maximized on $D_{val}$ for a model trained on $D_{p}^* \cup D_{tr}$.
Formally:
\begin{equation}
D_p^* \in \arg\max_{D_{p}} A(D_{val}, w^*) 
\label{eq:1}
\end{equation}
\begin{equation}
s.t. \quad w^* \in \arg\min_{w \in W} L(D_{tr} \cup D_p, w)
\label{eq:2}
\end{equation}

Problem~\ref{eq:1} is called `outer problem', since it depends on the model parameters $w^*$, which are found during model training (`inner problem', Equation~\ref{eq:2}).
Thus, finding a set of poison data $D^*_p$ which maximizes the attacker reward is a bi-level optimization problem.
Note that for DoS poisoning attacks, maximizing the attacker's objective function $A$ is equivalent to maximizing the model loss $L$, thus $A = L$ in Equation~\ref{eq:1}.

Every poison sample is initially assigned a class different from the original label.
The attack repeats poisoning steps, which modify the features of a malicious sample to increase its impact on the degradation of a model. Every poisoning step uses the concept of back-gradient optimization to derive $\nabla_{x_p}A$, which describes how the attacker's objective function varies w.r.t. a single poison instance $x_p$.
\begin{equation}
\nabla_{x_p} A = \nabla_{x_p} L + \frac{\partial w}{\partial x_p} \nabla_w L
\end{equation}
Computing $\frac{\partial w}{\partial x_p}$ to obtain how the solution of the learning algorithm changes w.r.t. the poison sample is a difficult task.
Early poisoning attacks have replaced the inner problem with its stationary KKT conditions, but this approach can be prohibitively expensive in time and memory, especially for larger neural networks (the attack has cubic complexity in the number of model parameters).
The approach presented in~\cite{munoz2017towards} replaces the inner optimization with a sequence of learning steps to smoothly update the parameters $w$. 
After an incomplete optimization of the inner problem within $T$ iterations, the parameters $w_T$ can be used to compute the desired gradients of the outer problem.
This significantly reduces the complexity of computing the outer gradient and enables to poison large neural networks and deep learning algorithms.

Pseudo-code algorithms in \cite{munoz2017towards} summarize how to derive $\nabla_{x_p} A$ by reversing a gradient-descent procedure with a fixed number of epochs and a fixed step size.
We refer to this attack as back-gradient optimization attack and use it to evaluate our defence against a strong, perfect-knowledge adversary.

We give a visual example of this attack in Figure \ref{fig:points_heat_map_loss} and \ref{fig:points_heat_map_accuracy}.
These figures depict a dataset (the \emph{points} dataset, see Section~\ref{ss:datasets}) with three classes (shown as blue, red and green dots) over a two dimensional feature space.
The black crosses represent a single poisoned data point over the course of 26 attack iterations.
This poison sample is created by taking a legitimate data instance, flipping its label to a poison label (from 2 to 1), and then iteratively moving it to a region where it will induce higher validation loss and lower validation accuracy.
The effects of training a model on the normal data $D_{tr}$ plus a single poisoned instance $x_p$ are shown by the background color map, whose $x_1$, $x_2$ coordinates represent the coordinates of the poisoned instance $x_p$.
To create the background plot, we partition the $x_1$, $x_2$ feature space into a grid, train a model for each poison instance on the grid, and plot validation loss and accuracy.
This gives us the `ground truth' with respect to how the poison sample should be changed in order to be maximally effective.
Observe how the attack modifies the point such that it is moved to regions of higher validation loss and lower validation accuracy.

\section{Poisoning Defence}\label{s:poisoning_defence}
While the previous section detailed strong baseline attacks from related work, this section presents our contribution:
A new defence to detect DoS data poisoning, which we later evaluate against the previously presented attacks, and which improves over previously presented defences.

\subsection{Requirements}\label{ss:requirements}
Our defence is designed to identify poison samples in a dataset without knowledge of the poison rate $\nu$. 
It requires only the model and the poisoned dataset, both of which is available to the defender.
This is the most unrestricted input conceivable, and shows the applicability of our defence in any real world setting.

\subsection{System Pipeline}\label{system_pipeline}
We present an algorithm that identifies poison instances and removes them from the dataset, while legitimate data remains. 
This approach is inspired by \cite{muller2019identifying}, where the authors detect mislabeled instances in classification datasets.

We define the following notation:
Let $D$ be a dataset $D=(\mathbf{x},\mathbf{y})$, consisting of $N$ instances out of $C$ classes. 
Input $\mathbf{x}$ represents $N$ data instances while target $\mathbf{y}$ represents $N$ data labels as one hot-encoded vectors of length $C$.
Both $\mathbf{x}$ and $\mathbf{y}$ can be conveniently expressed in matrix notation.
Finally, let $D^*_p \subseteq D$ for $D = (\mathbf{x},\mathbf{y})$ be the set of poison instances, which is unknown to the defender.

To identify these poison samples, we train a classifier $M$ on the given dataset $D$ and use the same classifier to obtain new class probabilities for $\mathbf{x}$. 
We then discard instances for which the class probability is small, which means that the classifier considers the original label extremely unlikely based on the feature distribution learned during training. 
This works because there is still a significant majority of unpoisoned data in the training dataset.
It is a reasonable assumption to make for crowd-sourced datasets, where a malicious individual can only introduce a small portion of malicious data.
Nevertheless, we show in our experiments that our system works with up to $10\%$ of poison data in the training dataset.

Formally, our proposed defence corresponds to an indicator function $f \to \{0, 1\}$ where $f(d) = 1$ indicates that $d \in D$ is poisoned.
We aim at maximizing the intersection between the true poisoned samples $D^*_p$ and $\{d \in D | f(d) = 1 \}$ while minimizing the intersection of $D^*_p$ and $\{d \in D | f(d) = 0 \}$.
Our proposed algorithm is as follows:
\begin{enumerate}
    \itemsep0em
    \item Prerequisites: Given a dataset $D=(\mathbf{x},\mathbf{y})$ with possibly crafted poison instances, i.e. $D = D_{tr} \cup D^*_p$. We allow $D^*_p$ to be empty (no attack is present).
    \item Train model $M$ on the dataset $D$.
    \item Reclassify training input $x$ using the trained model $M$ and obtain the new class probabilities (or alternatively logits) $M(\mathbf{x})=\mathbf{y'}$.
    \item Calculate for all $n\in \{0, ..., N-1\}$ the label likelihood $l_n$
    that instance $x_n$ is assigned the original label $y_n$, by computing $l_n := \langle y_n,y'_n \rangle = y'_{n_c}$. Here, $c$ is the non-zero valued index in the one-hot vector $y_n$.
    \item Sort all training instances according to $l_n$ in ascending order. This yields a figure as presented in Figure~\ref{fig:back_gradient_poisoning_kneeplot}, where instances with small likelihood w.r.t their original label are on the left end of the plot.
    \item Determine a threshold $k \in \{0, ..., N-1\}$ to distinguish between wrongly labeled poison data and correctly labeled benign data. 
    We find this threshold by approximating the first derivative of $l_n$ over a sliding window, and simply choosing $k$ as the argmax. We motivate this in Section~\ref{ss:defence}.
    \item Remove instances $D_p'$ where $l_i \leq l_k$ (i.e. left of the cutoff point) from the training set.
    Retrain the model on $D - D_p'$.
    \item Iteratively apply steps 2 to 7 until convergence.
\end{enumerate}

The intuition behind this algorithm is as follows:
By computing the likelihood $l_n$, we estimate the probability that the given label is the true label.
This requires a model $M$ which can generalize well even in the presence of adversarial data, such that it learns from the benign data instances what the poisoned instances' labels should be.
We then sort the instances by $l_n$ and remove a portion which has very small $l_n$ (and thus likely has a wrong label, which indicates adversarial data).
We retrain our model on the remaining data and iteratively re-evaluate the whole dataset.
In our empirical evaluation in Section~\ref{s:empircal_eval}, we find that this approach reliably identifies poisoned instances with low false positive and false negative rates.

Concerning the model $M$, we note that it must be able to generalize well and not overfit the dataset, so that deviations from the overall distribution of a class can be easily found.
Neural networks can satisfy these requirements. 
To prevent overfitting, we use an aggressive learning rate and early stopping.

With this system pipeline, we overcome fundamental restrictions of previous defences against data poisoning (see Section \ref{ss:poisoning_defence}).

\section{Evaluation}\label{s:empircal_eval}
In this section, we evaluate our defence against the attacks presented in Section~\ref{s:poisoning_attack}.
We find that these attacks, especially the back-gradient attack, considerably deteriorate the dataset, and that our defence can effectively mitigate this attack. Finally, we compare our defence against previously published data poisoning defences.


\subsection{Experimental Setup}\label{ss:datasets}
We evaluate our model on seven numerical and image datasets from keras \cite{lecun2010mnist, DBLP:journals/corr/abs-1708-07747} and the UCI Machine Learning Repository \cite{Dua:2019}, as presented in Table \ref{table:datasets}.
Poison instances $D_p$ are sampled from the validation set $D_{val}$ at run time.
Our main model is a neural network with one inner layer consisting of 10 neurons, as is appropriate for the small datasets we use.
This is in line with related work~\cite{munoz2017towards}.
For the image datasets, we also perform these experiments with a two-layer convolutional network.

\setlength{\tabcolsep}{4pt}
\begin{table}\centering
\caption{Datasets used in our empirical evaluation}
\label{table:datasets}
\begin{tabular}{@{}llcrrrr@{}}
\toprule
Name & {Size} & Classes & $|D_{tr}|$ &  $|D_{val}|$ & $|D_{test}|$\\
\midrule
breast\_cancer & (540, 30) & 2 & 100 & 220 & 220\\
fashion\_mnist\_156 & (6300, 784) & 3 & 300 & 3000 & 3000\\
fashion\_mnist\_17 & (2100, 784) & 2 & 100 & 1000 & 1000\\
mnist\_156 & (6300, 784) & 3 & 300 & 3000 & 3000\\
mnist\_17 & (2100, 784) & 2 & 100 & 1000 & 1000\\
points & (300, 2) & 3 & 100 & 100 & 100\\
spambase & (4200, 57) & 2 & 200 & 2000 & 2000\\
\bottomrule
\end{tabular}
\end{table}
\setlength{\tabcolsep}{1.4pt}

\subsection{Attack Results}\label{ss:attack}

\setlength{\tabcolsep}{3pt}
\begin{table}
\centering
\caption{Results of applying the label flipping and back-gradient attack against a neural network with one inner layer, averaged over 5 runs. Poisoning 10\% of the training data degrades the model's performance with respect to the test loss and test accuracy.
}
\label{tab:back_gradient_attack_results}
\begin{tabular}{@{}lccccccccc@{}}
\toprule
dataset & \multicolumn{3}{c}{test loss} &\multicolumn{3}{c}{test accuracy}\\
\cmidrule(lr{.50em}){2-4} \cmidrule(lr{.50em}){5-7}
& clean & flip & back-grad & clean & flip & back-grad\\
\midrule
breast\_cancer &             0.14 &               0.21 &          0.30 &                 0.96 &                   0.93 &              0.89 \\
f\_mnist\_156 &             0.09 &               0.24 &          0.30 &                 0.98 &                   0.92 &              0.92 \\
f\_mnist\_17 &             0.00 &               0.19 &          0.19 &                 1.00 &                   0.92 &              0.92 \\
mnist\_156 &             0.14 &               0.44 &          0.50 &                 0.96 &                   0.88 &              0.88 \\
mnist\_17 &             0.04 &               0.31 &          0.56 &                 0.99 &                   0.91 &              0.86 \\
points &             0.21 &               0.26 &          0.26 &                 0.94 &                   0.93 &              0.93 \\
spambase &             0.47 &               0.61 &          0.87 &                 0.88 &                   0.82 &              0.79 \\
\hline\hline
\textbf{mean} &             \textbf{0.16} &               \textbf{0.32} &          \textbf{0.42} &                 \textbf{0.96} &                   \textbf{0.90} &              \textbf{0.88} \\
\bottomrule
\end{tabular}
\end{table}


\begin{figure*}
\begin{subfigure}[b]{0.49\textwidth}
\centering
\includegraphics[width=\textwidth]{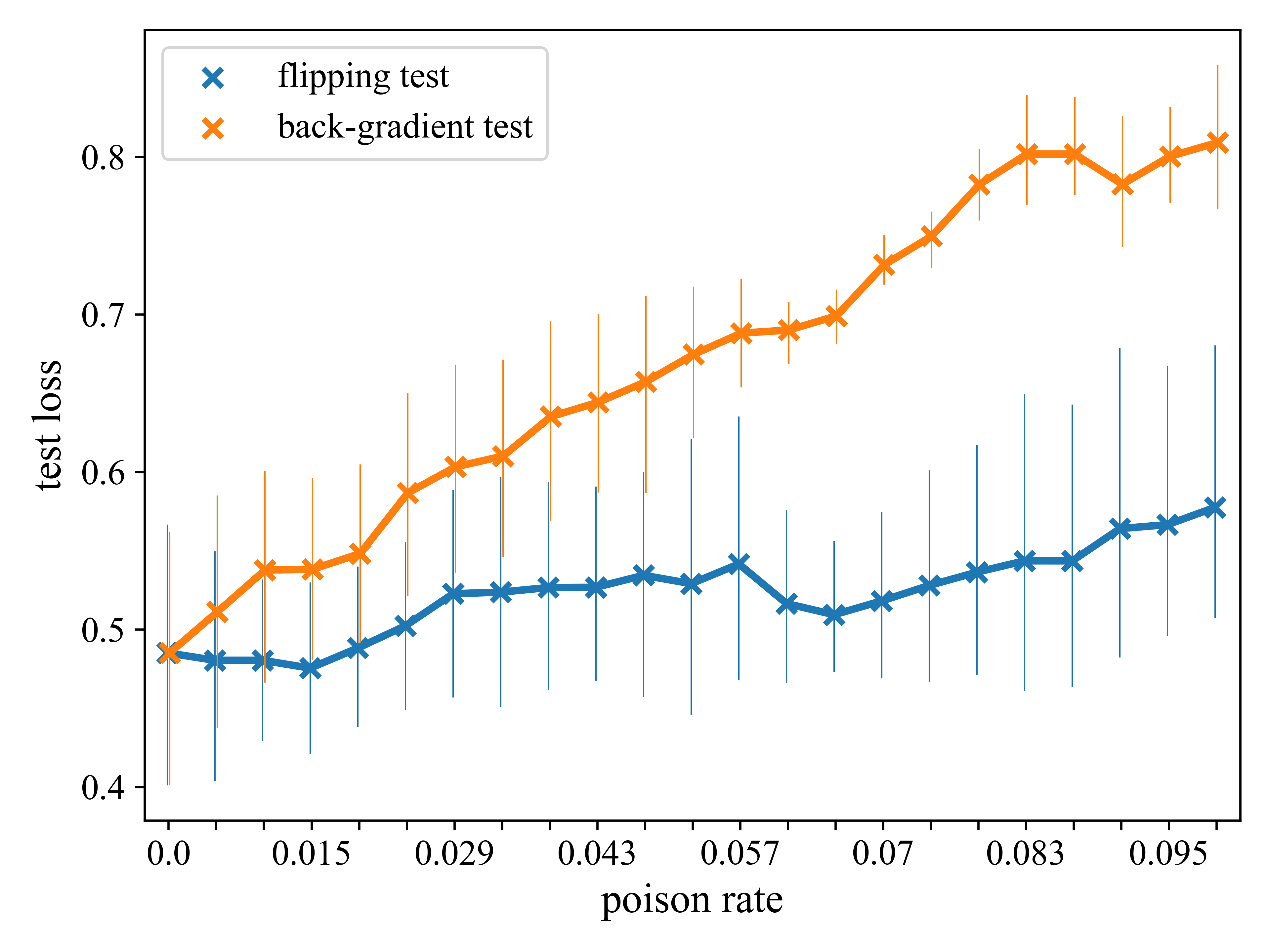}
\label{fig:car_loss}
\end{subfigure}
\hfill
\begin{subfigure}[b]{0.49\textwidth}
\centering
\includegraphics[width=\textwidth]{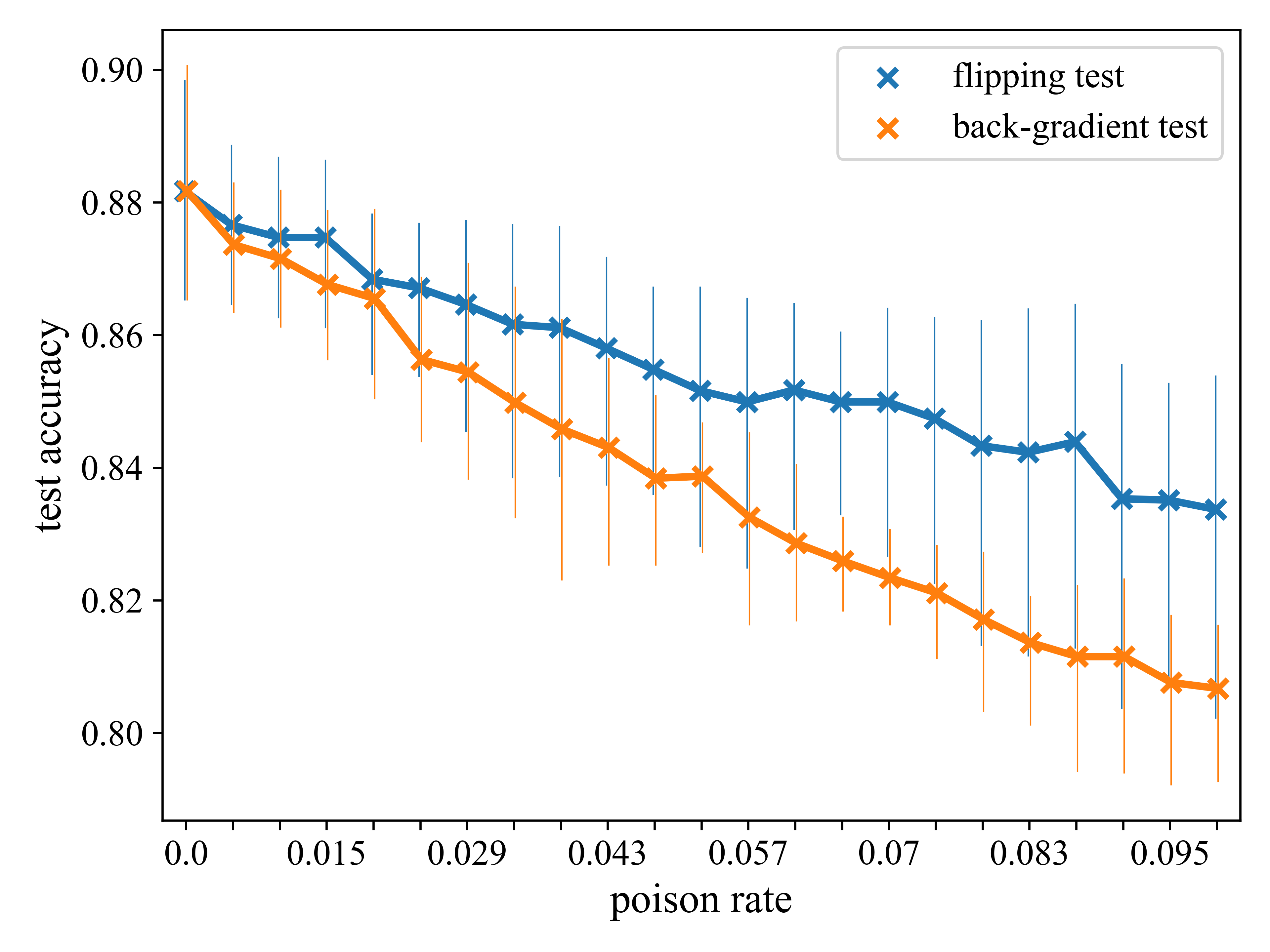}
\label{fig:car_accuracy}
\end{subfigure}
\caption{Degrading a model's performance by increasing the fraction of poison samples in the \emph{spambase} training dataset. The test loss and accuracy are averaged over 5 runs. Error bars represent the standard deviation.
}
\label{fig:car_loss_acc_cosine_all}
\end{figure*}

In this section, we present the results of the data poisoning attacks (c.f. Section~\ref{s:poisoning_attack}) on our datasets.
We inject up to $10\%$ poison data, and average all results of five individual runs.


The results for an exemplary dataset (\emph{spambase}) are shown in Figure~\ref{fig:car_loss_acc_cosine_all}.
An individual breakdown of the results can be found in Table~\ref{tab:back_gradient_attack_results}.
It can be seen that both attacks are effective, and that the 'back-gradient' attack consistently outperforms the more naive 'flipping' attack.

Additionally, we evaluate the effectiveness of the 'back-gradient' attack by plotting the model's decision surface.
Figure~\ref{fig:dec_borders_all} in the \emph{Appendix} shows a \emph{points} classification task with no adversarial data poisoning, as well as data poisoning induced by the flipping and back-gradient attack.
Observe how the attack alters the decision surface to possibly deteriorate the loss and accuracy.

Finally, we apply the 'back-gradient' optimization attack to a convolutional neural network consisting of two convolutional layers ($kernel\_size=(3,3)$, $pool\_size=(2,2)$).
As the authors in \cite{munoz2017towards} report, convolutional neural networks seem to be more resilient against optimal poisoning attacks.
We can confirm this result empirically, see Table  \ref{tab:back_gradient_defence_cnn} in the \emph{Appendix}.
 

\subsection{An Example of Applying our Defence}\label{ss:defence}
In this section, we apply our defence against a single, exemplary dataset to further illustrate how the defence works.

\begin{figure*}
    \centering
    \includegraphics[width=0.99\textwidth]{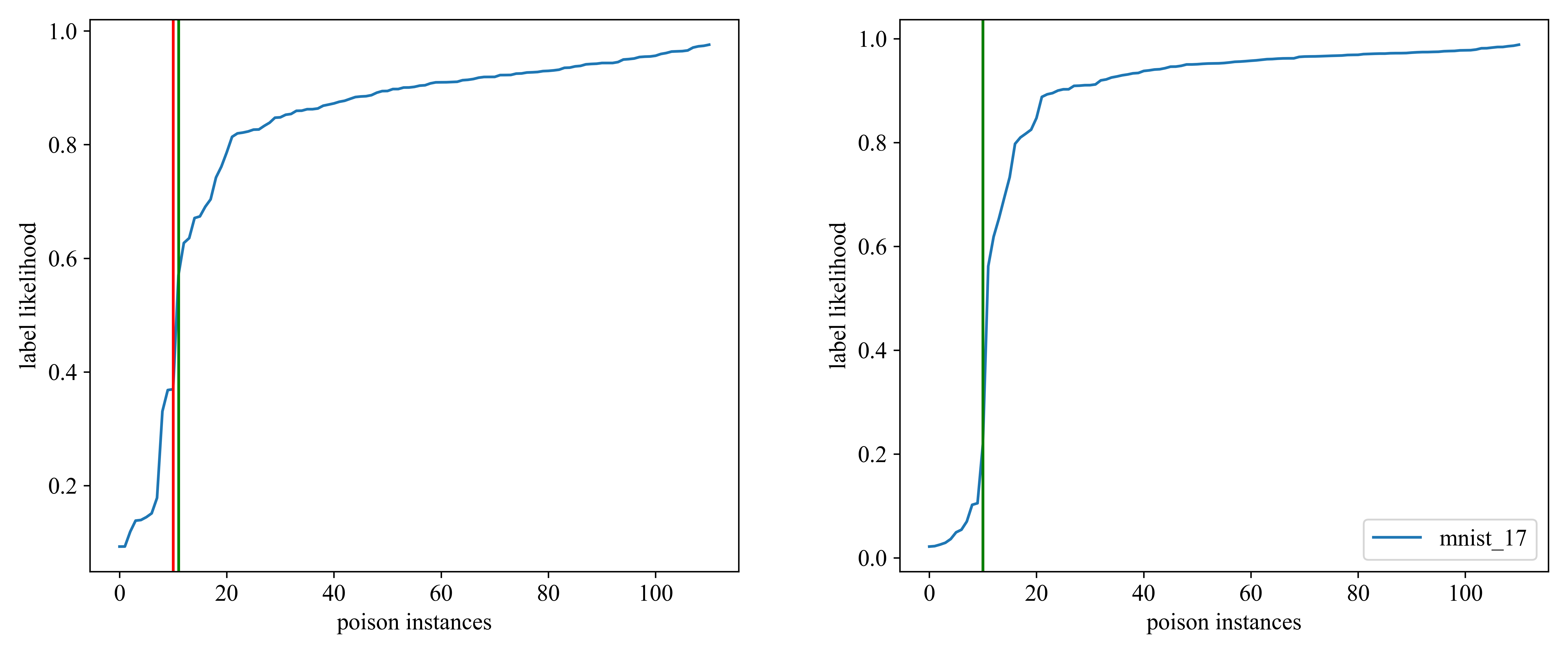}
    \caption{The result of applying our defense until convergence (here: two steps) on the \emph{mnist\_17} training dataset.
    Each image shows a `knee plots'  of label likelihood values $l_n$ (blue), sorted in ascending order.
    The x-axis lists the $l_n$-sorted instances, while the y-axis shows the corresponding label likelihood-values $l_n$.
    The red line represents the true fraction of poison samples (10\%), while the green line represents the fraction estimated by our defence algorithm.
    Our defence is accurate in estimating the poison rate (c.f. right image, where the red and green line overlap).
    Additionally, it identifies individual instances with a low $FPR=0.014$ and a low $FNR=0.013$.
    This results in the poisoned test loss being restored from $0.58$ to $0.10$, and the test accuracy being increased from $0.86$ to $0.97$.}
    \label{fig:back_gradient_poisoning_kneeplot}
\end{figure*}

\setlength{\tabcolsep}{5pt}
\begin{table*}
\centering
\caption{Results of our defence against the back-gradient optimization attack on a neural network with one inner layer, averaged over 5 runs. The datasets contain 10\% poison samples each. The column $\hat{\nu}$ shows the poison rate estimated by the defence algorithm while the next two columns compare FPR and FNR of our defence against a random baseline. 
Test loss and accuracy improve considerably when using the defence on the poisoned datasets.
}
\label{tab:back_gradient_defence_overview}
\begin{tabular}{@{}lccccccccc@{}}
\toprule
dataset &$\hat{\nu}$ & \multicolumn{2}{c}{FP} & \multicolumn{2}{c}{FN} & \multicolumn{2}{c}{test loss} & \multicolumn{2}{c}{test accuracy}\\
\cmidrule(lr{.50em}){3-4} \cmidrule(lr{.50em}){5-6} \cmidrule(lr{.50em}){7-8} \cmidrule(lr{.50em}){9-10}
 & & random & defence & random & defence & before & after & before & after\\
\midrule
breast\_cancer     &   0.088 &                        0.09 &                0.023 &                        0.09 &                0.034 &              0.32 &             0.31 &                  0.89 &                 0.92 \\
fashion\_mnist\_156 &   0.107 &                        0.09 &                0.016 &                        0.09 &                0.008 &              0.30 &             0.16 &                  0.92 &                 0.97 \\
fashion\_mnist\_17  &   0.095 &                        0.09 &                0.000 &                        0.09 &                0.004 &              0.18 &             0.00 &                  0.93 &                 1.00 \\
mnist\_156         &   0.113 &                        0.09 &                0.022 &                        0.09 &                0.008 &              0.51 &             0.17 &                  0.88 &                 0.96 \\
mnist\_17          &   0.101 &                        0.09 &                0.014 &                        0.09 &                0.013 &              0.58 &             0.10 &                  0.86 &                 0.97 \\
points            &   0.070 &                        0.09 &                0.005 &                        0.09 &                0.034 &              0.26 &             0.24 &                  0.92 &                 0.94 \\
spambase          &   0.072 &                        0.09 &                0.028 &                        0.09 &                0.055 &              0.80 &             0.79 &                  0.81 &                 0.84 \\
\hline\hline
\textbf{mean}              &   \textbf{0.092} &                        \textbf{0.09} &                \textbf{0.016} &                        \textbf{0.09} &                \textbf{0.022} &              \textbf{0.42} &             \textbf{0.25} &                  \textbf{0.89} &                 \textbf{0.94} \\
\textbf{std}               &   \textbf{0.015} &                        \textbf{0.00} &                \textbf{0.009} &                        \textbf{0.00} &                \textbf{0.018} &              \textbf{0.20} &             \textbf{0.24} &                  \textbf{0.04} &                 \textbf{0.05} \\
\bottomrule
\end{tabular}
\end{table*}

Assume a poisoned dataset $D$.
After applying Steps 2-4 from Section~\ref{system_pipeline}, we obtain a model $M$ trained on $D$. 
This model yields $l_n = \langle y_n, {y'}_n \rangle$ for all instances in the dataset $D$.
Remember that this label likelihood $l_n$ denotes a likelihood of the instance being assigned the original label.

Step 5 consists of sorting all training instances according to $l_n$.
This results in a so-called knee plot.
Figure~\ref{fig:back_gradient_poisoning_kneeplot} gives an example for the \emph{mnist\_17} dataset.
While the red line shows the true rate of poison samples in our training dataset ($\nu=10\%$), the green line is the model's estimate of the poison rate.
This estimate $\hat{\nu}$ is selected as the highest increase in $l_n$ over $n$.
We compute $\hat{\nu}$ using a sliding window over $l_n$ in order to approximate the first derivative of $l_n$, and choose the argmax (Step 6).
We then retrain the model on the data where $l_n$ falls above the threshold, and iterate until convergence (Step 7-8).

The intuition behind this approach is as follows:
The poison estimate $\hat{\nu}$ separates values of low and high label likelihood $l_n$. 
Low values indicate a mismatch between feature and target values, which may indicate poisoned data.
An oracle would assign benign data a label likelihood of $l_n = 1$ and poisoned data a likelihood of $l_n = 0$.
Thus, the obvious choice of threshold would be the point of discontinuity in $l_n$.
Since we need to work with an imperfect classifier instead of an oracle, the discontinuity is not as pronounced.
However, the largest derivative is a solid approximation which we find to work well.

\subsection{Defence Results}
We apply our defence to two scenarios: Data poisoned with the label flipping attack, and the back-gradient optimization attack.
Table \ref{tab:back_gradient_defence_overview} presents the results of our defence after applying the system pipeline twice.
We identify poisoned instances with an averaged FPR=0.02 and FNR=0.02, significantly outperforming a random baseline where both FPR, FNR = 0.09.
FP corresponds to the non-poison samples \emph{left} of the threshold that our classifier wrongly assumes to be poisoned and will be discarded from the training dataset.
FN correspond to the poison samples \emph{right} of the threshold that our classifier wrongly assumes to be benign and will remain in the training dataset.
Note that our defence does not have knowledge of the poison rate.
Table \ref{tab:back_gradient_defence_low_rates} in the \emph{Appendix} provides an evaluation of non-poisoned models to demonstrate the utility of our defence. In this case, our method only discards few samples of low label likelihood with a minimum effect on the test loss and accuracy of the model.

\begin{figure}
    \centering
    \includegraphics[width=0.55\textwidth]{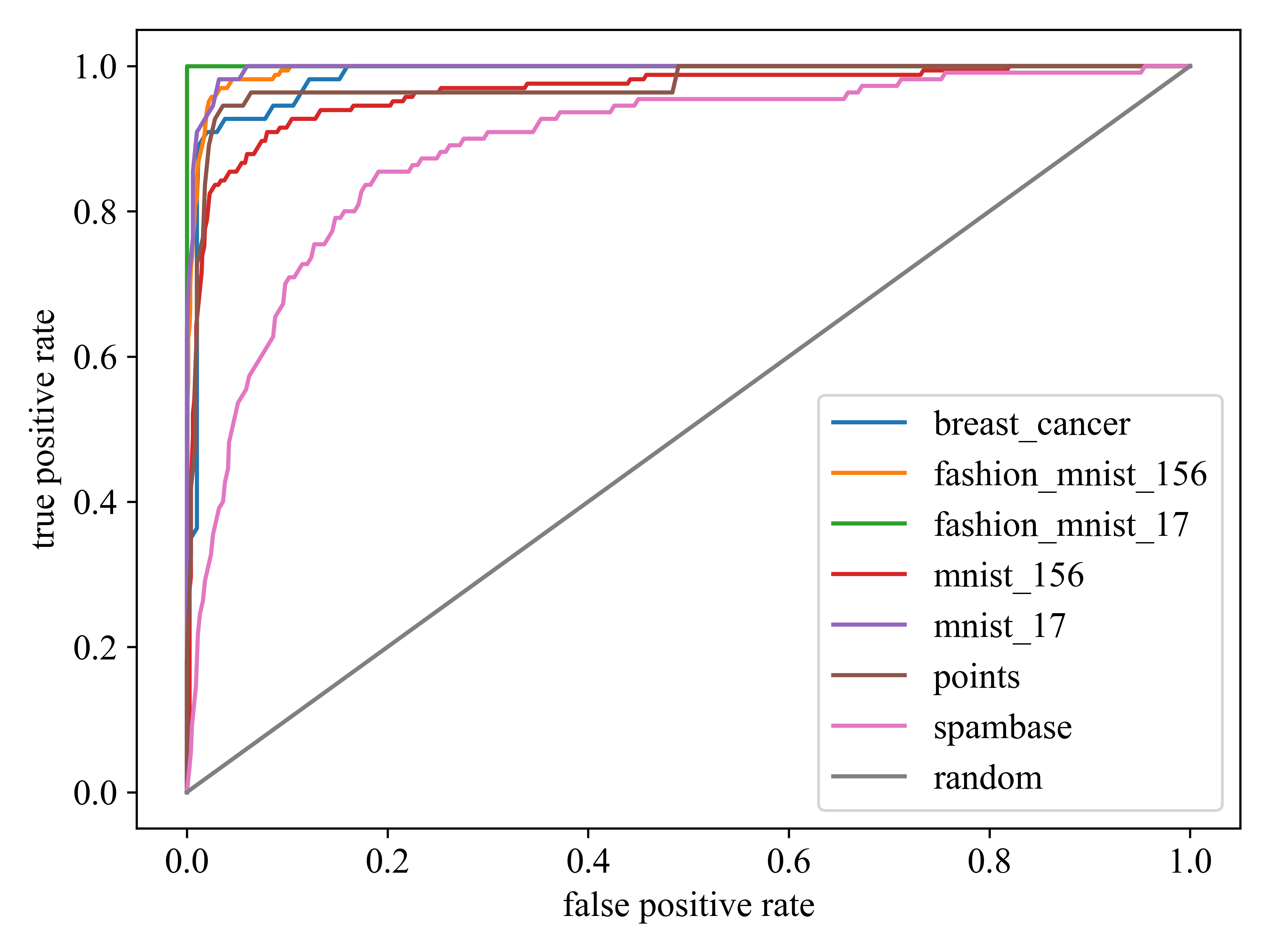}
    \caption{ROC curves averaged over 5 runs, obtained by applying our defence on seven datasets poisoned with the back-gradient optimization attack.
    The x-axis presents the fraction of false positive samples (benign data) while the y-axis presents the fraction of true positive samples (poison data) at varying  thresholds.
    }
    \label{fig:roc_curves}
\end{figure}

In order to show that this defence is not specifically tailored to any given learning model, we perform the same experiments with a convolutional network (instead of a dense network) and obtain similar results, c.f. Table~\ref{tab:back_gradient_defence_cnn} in the \emph{Appendix}.
ROC Curves are also supplied, c.f. Figure~\ref{fig:roc_curves}.


\setlength{\tabcolsep}{4pt}
\begin{table*}
\centering
\caption{Comparing our defence to outlier detection used in previous poisoning defences (see \cite{paudice2018detection, paudice2018label}), averaged over 5 runs.}
\label{tab:defence_comparison}
\begin{tabular}{@{}lcccccccc@{}}

\toprule
dataset & \multicolumn{4}{c}{FP} & \multicolumn{4}{c}{FN}\\
\cmidrule(lr{.50em}){2-5} \cmidrule(lr{.50em}){6-9}
 & \cite{paudice2018label} (kNN) & \cite{paudice2018detection} (L2)  & \cite{paudice2018detection} (LOF) & our defence & \cite{paudice2018label} (kNN) & \cite{paudice2018detection} (L2)  & \cite{paudice2018detection} (LOF) & our defence\\
\midrule
breast\_cancer     &                        0.034 &                        0.009 &                        0.031 &                0.023 &                        0.011 &                        0.085 &                        0.050 &                0.034 \\
fashion\_mnist\_156 &                        0.017 &                        0.059 &                        0.035 &                0.016 &                        0.009 &                        0.014 &                        0.032 &                0.008 \\
fashion\_mnist\_17  &                        0.000 &                        0.038 &                        0.040 &                0.000 &                        0.000 &                        0.000 &                        0.000 &                0.004 \\
mnist\_156         &                        0.019 &                        0.125 &                        0.046 &                0.022 &                        0.023 &                        0.012 &                        0.016 &                0.008 \\
mnist\_17          &                        0.025 &                        0.061 &                        0.034 &                0.014 &                        0.020 &                        0.032 &                        0.034 &                0.013 \\
points            &                        0.020 &                        0.025 &                        0.029 &                0.005 &                        0.013 &                        0.011 &                        0.013 &                0.034 \\
spambase          &                        0.098 &                        0.012 &                        0.042 &                0.028 &                        0.051 &                        0.097 &                        0.075 &                0.055 \\
\hline\hline
\textbf{mean}              &                        \textbf{0.031} &                        \textbf{0.047} &                        \textbf{0.037} &                \textbf{0.016} &                        \textbf{0.018} &                        \textbf{0.036} &                        \textbf{0.031} &                \textbf{0.022} \\
\textbf{std}               &                        \textbf{0.029} &                        \textbf{0.037} &                        \textbf{0.006} &                \textbf{0.009} &                        \textbf{0.015} &                        \textbf{0.036} &                        \textbf{0.023} &                \textbf{0.018}\\
\bottomrule
\end{tabular}
\end{table*}

\subsection{Evaluation Against Related Defences}
We implement two popular approaches for data-poisoning detection from related work and compare their results to our defence, c.f. Table \ref{tab:defence_comparison}.
The defence in \cite{paudice2018label} finds the $k$ nearest neighbours for each sample in the training set using the euclidean distance. 
If the fraction of data points with the most common label among these $k$ nearest neighbours is above a given threshold $t$ and if this most common label is different to the actual class of the sample, the sample is considered to be poisoned and will be rejected. 
Based on the experiments in \cite{paudice2018label}, we use $k=10$ and $t=0.6$ for all datasets.

In contrast, the outlier detection mechanism in \cite{paudice2018detection} first splits a small fraction of trusted data into the different classes and then trains an outlier detector for each class.
Every sample is assigned an outlierness score, which is either the euclidean distance (L2 norm) or the local outlier factor (LOF) with respect to its $k$ nearest neighbours.
The threshold to detect outliers is set by using the Empirical Cumulative Distribution Function of the benign training data and identifying the score at a certain $\alpha$-percentile. 
Samples with an outlierness score above that threshold are discarded. The parameters $k=5$ and $\alpha=0.99$, respectively $\alpha=0.95$, are chosen according to the author's suggestions.

We oberserve that all of these methods suffer from the curse of dimensionality, since they are based on a notion of metric distance.
Consequently, Table \ref{tab:defence_comparison} shows that our proposed defence obtains significantly lower FPR and FNR.
Experiments were conducted in real-world conditions:
We average over all datasets, and supply neither trusted reference data nor access to the ground-truth poisoning rate to our defence.

\section{Conclusion and Future Work}
In this paper, we present a new approach to defending against DoS poisoning attacks.
Our proposed mechanism can detect and discard malicious instances from a classification dataset, either by selecting a threshold automatically, or by sorting the instances for efficient triage by a human expert.
While previous work has mostly used clustering and outlier detection against poisoning attacks, we re-evaluate the training samples by the poisoned classifier.

The experimental evaluation on seven high-dimensional datasets has shown that label flipping and back-gradient optimization attacks significantly degrade the performance of neural networks. 
Applying our defence against datasets with $10\%$ of poison samples has demonstrated the reliable identification of malicious instances. Across all datasets our defence achieves a low $FPR=0.02$ and low $FNR=0.02$, consistently outperforming the random baseline ($FPR=FNR=0.09$).

We identify the following possibility for future work:
First, our system pipeline could be tested against more datasets and poisoning attacks. As the defence model learns from benign data instances what the poisoned instances' labels should be, our approach will most likely lack performance in the presence of unbalanced class sizes.
Second, our defence could be evaluated on a wider class of machine learning algorithms.
However, as long as the algorithm generalizes sufficiently, our approach should be transferable.

\newpage
\bibliographystyle{acm}
\bibliography{references}

\section*{Appendix}
\begin{figure}[h]
    \centering
    \begin{subfigure}[b]{0.36\textwidth}
    \includegraphics[width=\textwidth]{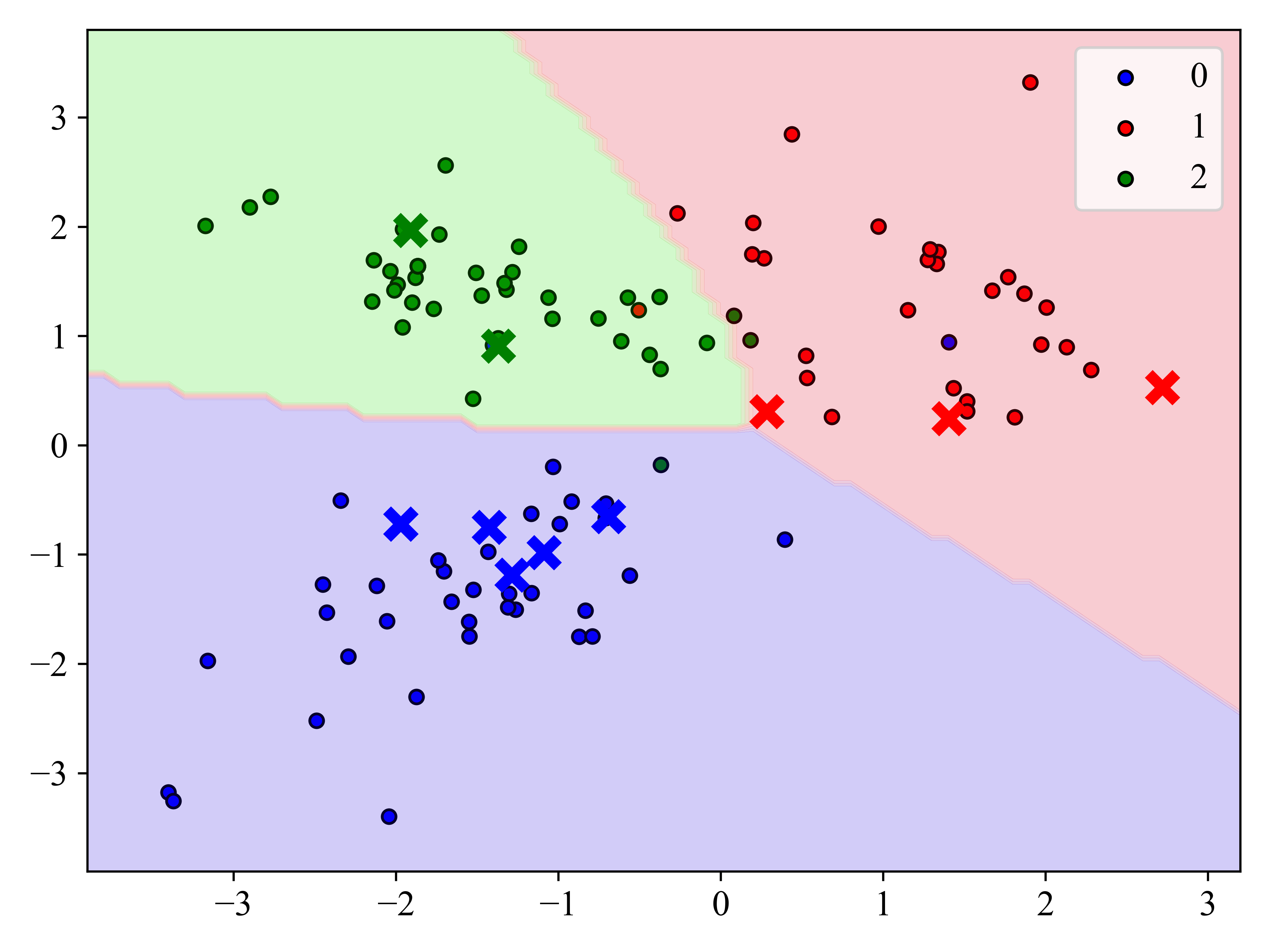}
    \caption{Before poisoning attack}
    \label{fig:decision_borders_before_poisoning}
  \end{subfigure}
    \begin{subfigure}[b]{0.36\textwidth}
    \includegraphics[width=\textwidth]{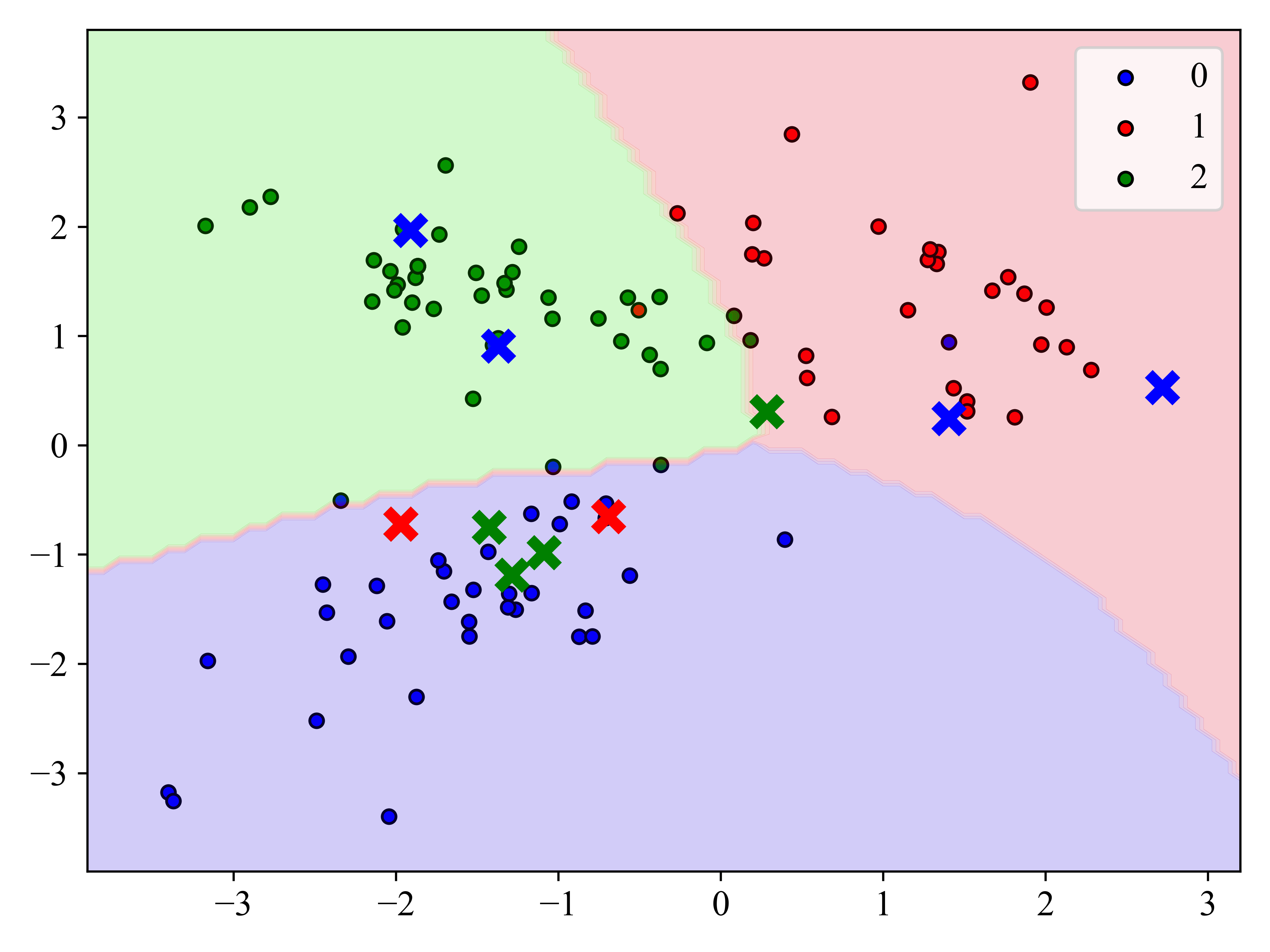}
    \caption{After label flipping}
    \label{fig:decision_borders_after_flipping}
    \end{subfigure}
    \begin{subfigure}[b]{0.36\textwidth}
    \includegraphics[width=\textwidth]{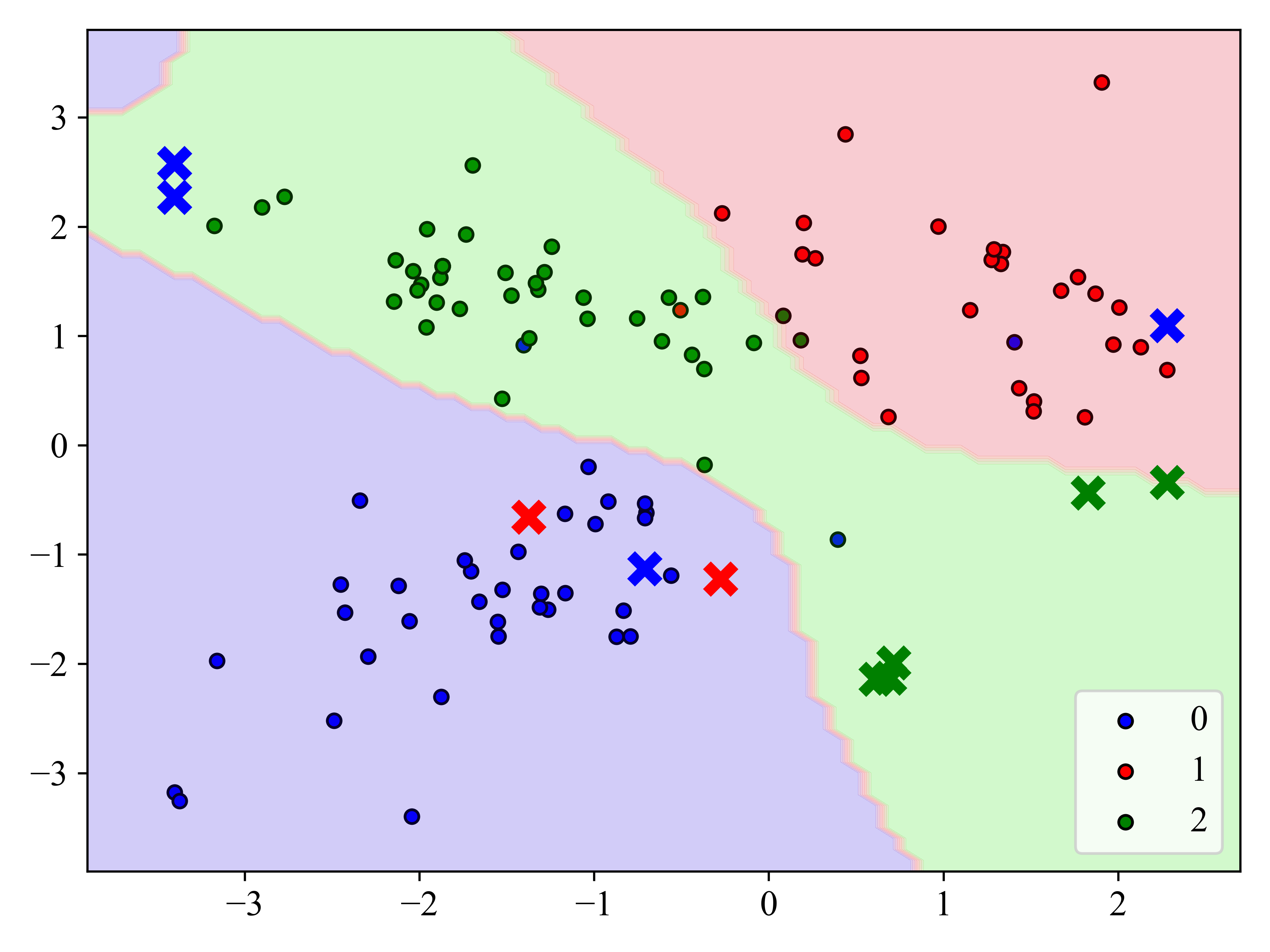}
    \caption{After back-gradient attack}
    \label{fig:decision_borders_after_poisoning}
  \end{subfigure}
  \caption{The effect of introducing 11 poison instances into a \emph{points} classification dataset. 
  The top figure shows the unpoisoned dataset, where the bold crosses indicate the data instances to be changed during the upcoming attacks.
  The middle figure shows the small change of the decision surface when the flip attack is applied.
  The bottom figure shows the considerably altered decision surface after the back-gradient attack.
  Note that not only the labels, but also the feature values of the poison data have changed.
  }
  \label{fig:dec_borders_all}
\end{figure}
\setlength{\tabcolsep}{5pt}
\begin{table*}
\centering
\caption{Results of our defence against the back-gradient optimization attack on a convolutional neural network, averaged over 5 runs. The datasets contain 10\% poison samples each. The column $\hat{\nu}$ shows the poison rate estimated by the defence algorithm while the next two columns compare FPR and FNR of our defence against a random baseline.
}
\label{tab:back_gradient_defence_cnn}
\begin{tabular}{@{}lccccccccc@{}}
\toprule
dataset &$\hat{\nu}$ & \multicolumn{2}{c}{FP} & \multicolumn{2}{c}{FN} & \multicolumn{2}{c}{test loss} & \multicolumn{2}{c}{test accuracy}\\
\cmidrule(lr{.50em}){3-4} \cmidrule(lr{.50em}){5-6} \cmidrule(lr{.50em}){7-8} \cmidrule(lr{.50em}){9-10}
 & & random & defence & random & defence & before & after & before & after\\
\midrule
fashion\_mnist\_156 &   0.125 &                        0.09 &                0.029 &                        0.09 &                0.002 &              0.17 &             0.17 &                  0.94 &                 0.97 \\
fashion\_mnist\_17  &   0.097 &                        0.09 &                0.000 &                        0.09 &                0.003 &              0.13 &             0.01 &                  0.99 &                 1.00 \\
mnist\_156         &   0.117 &                        0.09 &                0.025 &                        0.09 &                0.007 &              0.39 &             0.22 &                  0.86 &                 0.95 \\
mnist\_17          &   0.100 &                        0.09 &                0.006 &                        0.09 &                0.005 &              0.16 &             0.08 &                  0.95 &                 0.98 \\
\hline\hline
\textbf{mean}              &   \textbf{0.110} &                        \textbf{0.09} &                \textbf{0.015} &                        \textbf{0.09} &                \textbf{0.004} &              \textbf{0.21} &             \textbf{0.12} &                  \textbf{0.94} &                 \textbf{0.98} \\
\textbf{std}               &   \textbf{0.012} &                        \textbf{0.00} &                \textbf{0.012} &                        \textbf{0.00} &                \textbf{0.002} &              \textbf{0.10} &             \textbf{0.08} &                  \textbf{0.05} &                 \textbf{0.02} \\
\bottomrule
\end{tabular}
\end{table*}

\setlength{\tabcolsep}{5pt}
\begin{table*}
\centering
\caption{Results of our defence against the back-gradient optimization attack on a neural network with one inner layer, averaged over 5 runs. The datasets contain 0\% or 5\% poison samples each. The column $\hat{\nu}$ shows the poison rate estimated by the defence algorithm while the next two columns compare FPR and FNR of our defence against a random baseline. 
Test loss and accuracy remain consistent in the case of no poisoning or improve considerably when using the defence on the poisoned datasets.
}
\label{tab:back_gradient_defence_low_rates}
\begin{tabular}{@{}lcccccccccc@{}}
\toprule
dataset & $\nu$ & $\hat{\nu}$ & \multicolumn{2}{c}{FP} & \multicolumn{2}{c}{FN} & \multicolumn{2}{c}{test loss} & \multicolumn{2}{c}{test accuracy}\\
\cmidrule(lr{.50em}){4-5} \cmidrule(lr{.50em}){6-7} \cmidrule(lr{.50em}){8-9} \cmidrule(lr{.50em}){10-11}
 & & & random & defence & random & defence & before & after & before & after\\
\midrule
breast\_cancer     &         0.00 &   0.028 &                       0.000 &                0.028 &                       0.000 &                0.000 &              0.16 &             0.18 &                  0.96 &                 0.95 \\
fashion\_mnist\_156 &         0.00 &   0.010 &                       0.000 &                0.010 &                       0.000 &                0.000 &              0.09 &             0.11 &                  0.98 &                 0.98 \\
fashion\_mnist\_17  &         0.00 &   0.010 &                       0.000 &                0.010 &                       0.000 &                0.000 &              0.00 &             0.00 &                  1.00 &                 1.00 \\
mnist\_156         &         0.00 &   0.009 &                       0.000 &                0.009 &                       0.000 &                0.000 &              0.13 &             0.14 &                  0.96 &                 0.96 \\
mnist\_17          &         0.00 &   0.028 &                       0.000 &                0.028 &                       0.000 &                0.000 &              0.05 &             0.08 &                  0.98 &                 0.97 \\
points            &         0.00 &   0.024 &                       0.000 &                0.024 &                       0.000 &                0.000 &              0.20 &             0.23 &                  0.94 &                 0.94 \\
spambase          &         0.00 &   0.025 &                       0.000 &                0.025 &                       0.000 &                0.000 &              0.48 &             0.53 &                  0.88 &                 0.88 \\
\hline
\hline
\textbf{mean}              &         0.00 &   0.019 &                       0.000 &                0.019 &                       0.000 &                0.000 &              0.16 &             0.18 &                  0.96 &                 0.95 \\
\textbf{std}               &         0.00 &   0.017 &                       0.000 &                0.017 &                       0.000 &                0.000 &              0.16 &             0.17 &                  0.04 &                 0.04 \\
\midrule
breast\_cancer     &         0.05 &   0.071 &                       0.048 &                0.027 &                       0.057 &                0.004 &              0.30 &             0.25 &                  0.91 &                 0.95 \\
fashion\_mnist\_156 &         0.05 &   0.066 &                       0.048 &                0.022 &                       0.048 &                0.004 &              0.19 &             0.15 &                  0.95 &                 0.97 \\
fashion\_mnist\_17  &         0.05 &   0.046 &                       0.048 &                0.000 &                       0.057 &                0.002 &              0.07 &             0.00 &                  0.97 &                 1.00 \\
mnist\_156         &         0.05 &   0.053 &                       0.047 &                0.014 &                       0.047 &                0.009 &              0.36 &             0.18 &                  0.91 &                 0.95 \\
mnist\_17          &         0.05 &   0.057 &                       0.046 &                0.012 &                       0.055 &                0.002 &              0.33 &             0.06 &                  0.91 &                 0.98 \\
points            &         0.05 &   0.037 &                       0.048 &                0.006 &                       0.057 &                0.017 &              0.22 &             0.24 &                  0.93 &                 0.94 \\
spambase          &         0.05 &   0.069 &                       0.050 &                0.043 &                       0.050 &                0.021 &              0.68 &             0.59 &                  0.84 &                 0.88 \\
\hline
\hline
\textbf{mean}              &         0.05 &   0.057 &                       0.048 &                0.018 &                       0.053 &                0.008 &              0.31 &             0.21 &                  0.92 &                 0.95 \\
\textbf{std}               &         0.00 &   0.021 &                       0.003 &                0.019 &                       0.005 &                0.010 &              0.19 &             0.19 &                  0.04 &                 0.04 \\
\bottomrule
\end{tabular}
\end{table*}

\end{document}